# Quantum Adjoint Convolutional Layers for Effective Data Representation

Ren-Xin Zhao, Shi Wang*, and Yaonan Wang,

*Abstract*—Quantum Convolutional Layer (QCL) is considered as one of the core of Quantum Convolutional Neural Networks (QCNNs) due to its efficient data feature extraction capability. However, the current principle of QCL is not as mathematically understandable as Classical Convolutional Layer (CCL) due to its black-box structure. Moreover, classical data mapping in many QCLs is inefficient. To this end, firstly, the Quantum Adjoint Convolution Operation (QACO) consisting of a quantum amplitude encoding and its inverse is theoretically shown to be equivalent to the quantum normalization of the convolution operation based on the Frobenius inner product while achieving an efficient characterization of the data. Subsequently, QACO is extended into a Quantum Adjoint Convolutional Layer (QACL) by Quantum Phase Estimation (QPE) to compute all Frobenius inner products in parallel. At last, comparative simulation experiments are carried out on PennyLane and TensorFlow platforms, mainly for the two cases of kernel fixed and unfixed in QACL. The results demonstrate that QACL with the insight of special quantum properties for the same images, provides higher training accuracy in MNIST and Fashion MNIST classification experiments, but sacrifices the learning performance to some extent. Predictably, our research lays the foundation for the development of efficient and interpretable quantum convolutional networks and also advances the field of quantum machine vision.

## I. Introduction

**Q**CNNs are a new machine learning model that has emerged in recent years with the development of quantum computers [1–4]. The origin of QCNNs probably dates back to 2017, initially existing as theoretical explorations devoid of ansatz models[5]. Until 2019, a QCNN with multiscale entanglement reformulation ansatz as a convolutional layer was proposed, laying the foundation for its implementation on near-term quantum devices [6]. Subsequently, QCNNs exemplified by quantum random circuits [7], controlled rotating quantum gates [8], multiple CNOT gates [9], etc. as convolutional layers have been successively presented in pursuit of diversification of convolutional layers, which greatly enriched the field of QCNNs. Despite being in the early stages of research, QCNNs have demonstrated advantages in diverse fields, including image processing [10–12], physics analysis [13–15], and medical detection [16, 17]. Illustratively, QCNN exhibits a swifter learning rate and superior test accuracy compared to Classical Convolutional Neural Network (CCNN) with analogous parameters [18]. Furthermore, QCNNs pose challenges for classical modeling in certain aspects, making them apt for capturing more intricate features [19]. In fact, the reasons behind these advantages are closely tied to the ansatz design of QCNNs [20, 21], including the QCLs under consideration in this paper.

QCLs are important as the centerpiece of the research in this paper because they inherit the two advantages of CCLs, namely local area connectivity and weight sharing, which are the special features that distinguish convolutional neural networks from other networks. Local area connectivity and weight sharing decrease the number of weights, model complexity, and risk of overfitting, making the network easy to optimize [22, 23]. However, comparing Refs. [5–11, 13, 16, 17] with classical convolution operations such as dilated convolutions [24], transposed convolutions [25], the following problems are found.

1. Ansatzes with black-box structure, such as multiscale entanglement renormalization ansatz [7], alternating ansatz [10, 11], render the operating principle of quantum convolution difficult to grasp and understand due to the lack of interpretability. The benefit of developing interpretable QCLs is to facilitate its improvement, validation, and debugging, thus enhancing its credibility [26].

2. Inefficiencies in the data representation of QCL are likely to lead to failure in the training of quantum machine learning models.

In order to address the above two issues, the following main contributions are made in this paper, using a typical CCL based on Frobenius inner product as a research background.

- QACO, an interpretable quantum convolution operation capable of compressing classical data exponentially, is theoretically equivalent to the Frobenius inner product of two quantum amplitude encodings.
- A QACO ansatz is designed and extended to the QACL ansatz using QPE [30] for deducing all convolutional results at once.
- Experiments elucidate that models with QACL exhibit elevated training accuracy in the context of the MNIST and Fashion MNIST datasets. However, this enhanced accuracy comes at the cost of compromised learning performance.

Therefore this paper is organized as follows. In Section 2, some of the necessary theoretical foundations are briefly

Ren-Xin Zhao is with the School of Computer Science and Engineering, Central South Univerisity, China, Changsha, 410083. Shi Wang and Yaonan Wang are with the College of Electrical and Information Engineering, Hunan University, China, Changsha, 410082
Corresponding author: Shi Wang
E-mails: renxin_zhao@alu.hdu.edu.cn, shi_wang@hnu.edu.cn, yaonan@hnu.edu.cn

reviewed. In Section 3, the principle of interpretable QACO with efficient data representation properties is theoretically elaborated and the ansatz structure of QACO and QACL is proposed. In Section 4, the feasibility and effectiveness of the scheme of this paper are verified by three experiments. Finally, a conclusion is drawn.

## II. Preliminaries

In this section, the foundational tenets of classical convolutional operations, predicated on the Frobenius inner product—a metric of similarity, are initially expounded. Subsequently, methodologies for gauging similarity between quantum states in quantum computing are scrutinized. Finally, the review encompasses QPE, a pivotal technique instrumental in the amplification of QACO.

### A. Classical Convolution Operation based on Frobenius Inner Product

A square matrix

$$Map := \begin{bmatrix} a_{0,0} & \cdots & a_{0,n-1} \\ \vdots & \ddots & \vdots \\ a_{n-1,0} & \cdots & a_{n-1,n-1} \end{bmatrix} \quad (1)$$

of size $n \times n$ is defined. Meanwhile, there is a kernel (or filter)

$$Win := \begin{bmatrix} w_{0,0} & \cdots & w_{0,m-1} \\ \vdots & \ddots & \vdots \\ w_{l-1,0} & \cdots & w_{l-1,m-1} \end{bmatrix} \quad (2)$$

of size $l \times m$. Let a submatrix of Eq. (1) of the same size as Eq. (2) be

$$Map_{j,k} := \begin{bmatrix} a_{j,k} & \cdots & a_{j,k+m-1} \\ \vdots & \ddots & \vdots \\ a_{j+l-1,k} & \cdots & a_{j+l-1,k+m-1} \end{bmatrix}. \quad (3)$$

Then the matrix after a basic convolution operation becomes

$$Conv := \begin{bmatrix} F_{0,0} & \cdots & F_{0,n-m+1} \\ \vdots & \ddots & \vdots \\ F_{n-l+1,0} & \cdots & F_{n-l+1,n-m+1} \end{bmatrix}, \quad (4)$$

where

$$F_{j,k} := \langle Win, Map_{j,k} \rangle_F = \sum_a \sum_b a_{j+a,k+b} w_{a,b} \quad (5)$$

denotes the Frobenius inner product between Eq. (2) and Eq. (3). Obviously, Eq. (5) can be regarded as a measure of the similarity between two matrices. In the next subsection, methods for measuring the similarity between two quantum states in quantum computation are shown.

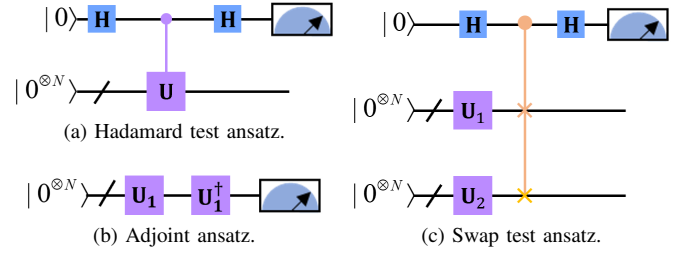

Fig. 1: Methods for overlap estimation.

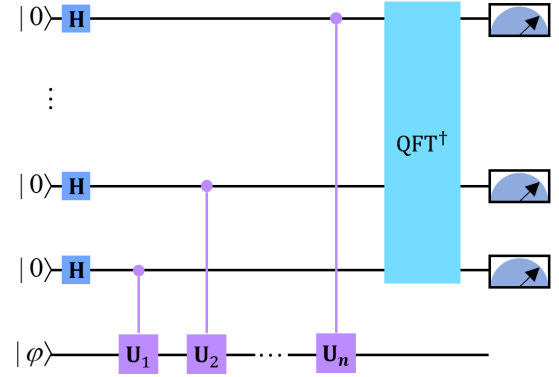

Fig. 2: Quantum phase estimation circuit.

### B. Quantum Overlap

In quantum computing, the similarity between quantum states is terminologically known as quantum overlap. Currently, there are three prevailing approaches for estimating quantum overlap, namely the adjoint method [27], the Hadamard test [28] and the swap test [29], as shown in Fig. 1. The Hadamard test ansatz composed of an auxiliary qubit, an input register $|0^{\otimes N}\rangle$, two Hadamard gates, and a controlled ansatz is depicted in Fig. 1a, where $\otimes$ is the tensor product symbol. $N$ is the number of qubits. The swap test ansatz in Fig. 1c is somewhat akin to that of Fig. 1a and consists of an auxiliary qubit, two input registers $|0^{\otimes N}\rangle$, two Hadamard gates, two ansatzes, and a number of controlled swap gates. The adjoint ansatz in Fig. 1b is made up of the input register $U_1$ and its inverse circuit $U_1^\dagger$. Notably, here, $U_1$ and its inverse circuit $U_1^\dagger$ are just structured as conjugate transpositions of each other, but their internal parameters can be distinct. With the above structural basis, the quantum overlaps of these three methods are

$$\text{Re}(\langle 0^{\otimes N}|U|0^{\otimes N}\rangle) = 2P(|0\rangle) - 1 \quad (6)$$

for Hadamard test,

$$||\langle 0^{\otimes N}|U_1 U_2^\dagger|0^{\otimes N}\rangle||^2 = 2P(|0\rangle) - 1 \quad (7)$$

for swap test and

$$||\langle 0^{\otimes N}|U_1 U_1^\dagger|0^{\otimes N}\rangle||^2 = P(|0\rangle) \quad (8)$$

for adjoint method, respectively, where $P(|0\rangle)$ is the probability of measuring the ground state $|0\rangle$. In conclusion, according to Eq. (6) to Eq. (8), unlike the adjoint method, the Hadamard test and the swap test are indirect ways to get the quantum overlap. Moreover, the quantum overlap form of the Hadamard test is different from that of the swap test and the adjoint method. The quantum overlap of Hadamard test can be negative, while the other two methods obviously cannot do it. Finally, if two quantum states of the same dimension are compared, it is clear that the adjoint method and the Hadamard test are more qubit efficient compared to the swap test, while the adjoint method has a deeper quantum circuit.

### C. Quantum Phase Estimation

QPE [30] shown in Fig. 2 calculates the phase of the eigenvalues of a given unitary operator $U_P$, i.e., solving for $\theta$ in $U_P|\varphi\rangle = e^{2\pi i\theta}|\varphi\rangle$, where $|\varphi\rangle$ is the eigenvector of $U_P$. From Fig. 2, it can be seen that the QPE consists of three parts. Assuming that the feature vector $|\varphi\rangle$ has been prepared, the QPE contains the following main steps: (1) Put the auxiliary quantum register in the superposition state. (2) Transfer the eigenvalue phase decomposition of $U\_P$ to the amplitude of the auxiliary qubits through a series of special controlled rotation quantum gates $C-U_1,\cdots,C-U_n$. (3) Perform inverse quantum Fourier transform $QFT^\dagger$ on the auxiliary qubits to phase shift the eigenvalues on the amplitude to the basis vector. (4) At last, the phase information of the eigenvalues can be acquired by synthesizing the base vectors of the auxiliary qubits after measuring them separately.

## III. QUANTUM ADJOINT CONVOLUTIONAL LAYER FOR EFFICIENT DATA REPRESENTATION

In this section, our **contributions** include: (1) An interpretable QACO with exponential compression of classical data is proposed, which is theoretically equivalent to the Frobenius inner product of two amplitude encodings. (2) The ansatz structure of QACO and QACL is designed.

### A. Quantum Adjoint Convolution Operation

Efficient data characterization entails the capacity not only to exponentially compress classical data but also to maintain transparency and interpretability in principle. Our first **innovation** is to propose QACO

$$\langle \Phi_{j,k}|\Psi\rangle = \sum_{t=0}^{2^N-1} \alpha_t \beta_t. \quad (9)$$

to realize this effective expression. QACO is a quantum version of Eq. (5), where $\langle\Phi_{j,k}|$ and $|\Psi\rangle$ correspond to Eq. (3) and Eq. (2), respectively. $N = \lceil \log_2(l \times m) \rceil$ is the count of qubits. $\alpha_t$ and $\beta_t$ indicate amplitude values on the $t$-th basis vector.

Proceeding further, the scrutiny of the derivation of Eq. (9) commences. First, the elements from Eq. (3) can be arranged in rows to obtain the sequence

$$\{\{a_{j+c,k+d}\}_{c=0}^{m-1}\}_{d=0}^{l-1} \in \mathcal{X}, \quad (10)$$

where $\mathcal{X}$ represents the classical data space. Then each data point in Eq. (18) is quantum normalized so as to obtain a new sequence

$$\{\{\mathcal{A}_{c,d}^{j,k}\}_{c=0}^{m-1}\}_{d=0}^{l-1} = \left\{ \left\{ \frac{a_{j+c,k+d}}{\sqrt{\Sigma_a^2}} \right\}_{c=0}^{m-1} \right\}_{d=0}^{l-1} \in \mathcal{H}, \quad (11)$$

where $\Sigma_a^2 = \sum_{c=0}^{m-1}\sum_{d=0}^{l-1} a_{j+c,k+d}^2$. $\mathcal{H}$ denotes the Hilbert space. Now, each $\mathcal{A}_{c,d}^{j,k}$ in Eq. (11) can be embedded into the quantum system to yield the quantum state

$$|\Phi_{j,k}\rangle = \sum_{c=0}^{m-1}\sum_{d=0}^{l-1} \mathcal{A}_{c,d}^{j,k}|c+d\rangle + \sum_{i=m\times l}^{2^N-1} 0|i\rangle \in \mathcal{H}, \quad (12)$$

where $|c+d\rangle$ and $|i\rangle$ are a set of standard orthogonal bases in Hilbert space. In this way, Eq. (12) establishes a precise correspondence between the serial number of the standard orthogonal basis and its preceding amplitude value with the position and element values in sequence (10) derived from the image block conversion.

Here, for ease of interpretation, Eq. (12) is simplified to

$$|\Phi_{j,k}\rangle = \sum_{t=0}^{2^N-1} \alpha_t|t\rangle, \quad (13)$$

where

$$\alpha_t = \begin{cases} \mathcal{A}_{c,d}^{j,k}, & t = c+d \in [0, l\times m-1] \\ 0, & else \end{cases} \quad (14)$$

is the amplitude value on the $t$-th basis vector. It can be seen that Eq. (12) converts classical data into amplitude values on standard orthogonal bases, which is called quantum amplitude encoding $Amp(\theta)$ [31]. This encoding can represent up to $2^N$ classical data points in $N$ qubits. In other words, it stores an exponential amount of classical information by sacrificing a small number of qubits as therefore an efficient and compact way of encoding. Furthermore, as mentioned above, this encoding creates an interpretable correspondence with the image elements and their locations.

Similarly, the efficient quantum representation of Eq. (2) is

$$|\Psi\rangle = \sum_{t=0}^{2^N-1} \beta_t|t\rangle, \quad (15)$$

where

$$\beta_t = \begin{cases} \mathcal{W}_{c,d}, & t = c+d \in [0, l\times m-1] \\ 0, & else \end{cases}. \quad (16)$$

$\mathcal{W}_{c,d}$ in Eq. (16) is from sequence

$$\{\{\mathcal{W}_{c,d}\}_{c=0}^{m-1}\}_{d=0}^{l-1} = \left\{ \left\{ \frac{w_{c,d}}{\sqrt{\Sigma_w^2}} \right\}_{c=0}^{m-1} \right\}_{d=0}^{l-1} \in \mathcal{H} \quad (17)$$

with $\Sigma_w^2 = \sum_{c=0}^{m-1}\sum_{d=0}^{l-1} w_{c,d}^2$, which is the consequence of the quantum normalization of the sequence

$$\{\{w_{j+c,k+d}\}_{c=0}^{m-1}\}_{d=0}^{l-1} \in \mathcal{X} \quad (18)$$

from Eq. (2).

Combining Eq. (13) and Eq. (15), the overlap between quantum states can be described as follows:

$$\langle\Phi_{j,k}|\Psi\rangle = \sum_{a=0}^{2^N-1}\sum_{b=0}^{2^N-1} \alpha_a \beta_b \langle a|b\rangle. \quad (19)$$

Eq. (19) illustrates that as long as the bases are orthogonal, i.e., $|a\rangle \neq |b\rangle$, then the overlap of quantum states $\langle a|b\rangle$ is 0, otherwise the overlap of quantum states $\langle a|b\rangle$ is 1. In a nutshell,

Eq. (19) only needs to focus on the sum of the products of $\alpha_a$ and $\beta_a$ when $|a\rangle = |b\rangle$, which matches the operation of Eq. (5) very well. Hence, Eq. (19) can be rewritten as Eq. (9). Up to this point, the physical meaning of Eq. (9) has corresponded to Eq. (5), which means that the Frobenius inner product can be solved by quantum overlap. The next subsection specifies how to implement Eq. (9) through quantum circuits.

### B. Quantum Adjoint Convolutional Layer

This subsection focus on practical applications to create specific quantum models for the theory of the previous subsection for future implementation on quantum computers.

*1) Ansatz Structure of Quantum Adjoint Convolutional Operation:* First, comparing Eq. (6) to Eq. (8), Eq. (6) can characterize positive and negative values, thus the Hadamard test realizes QACO in Eq. (9) more realistically. For this reason, modified by Fig. 1a, **QACO ansatz**

$$(H \otimes I^{\otimes N})C - Amp^\dagger(\theta') \times \\ C - Amp(\theta)(H \otimes I^{\otimes N})|0^{\otimes(N+1)}\rangle, \tag{20}$$

in the blue box in Fig. 3 is designed by us, Where $H$ is the Hadamard gate. $C - Amp^\dagger(\theta')$ and $C - Amp(\theta)$ are two controlled quantum amplitude encoding ansatzes. Furthermore, it is important to distinguish that $C-Amp(\theta)$ (or $C-Amp^\dagger(\theta')$) is applied exclusively when the auxiliary qubit is $|1\rangle$ (or $|0\rangle$). $I$ denotes the identity matrix. Structurally, this QACO with one more controlled ansatz than Fig. 1a, effectively estimates the Frobenius inner product between the cyan region in an image and the kernel located in the orange region. An additional controlled ansatz is introduced here since the cyan controlled ansatz $C-Amp(\theta)$ only depicts local information in an image, while an orange one $C-Amp^\dagger(\theta')$ is needed to load the kernel information. Ultimately, the measurements of QACO ansatz are stored in the green region depicted in Fig. 3. Specifically, the workflow of QACO anatz is as follows.

Step 1: Initialize the auxiliary register and input register:

$$|\varphi\rangle = |0\rangle \otimes |0^{\otimes N}\rangle. \tag{21}$$

Step 2: Pass through the Hadamar gate:

$$|\varphi\rangle \xrightarrow{H \otimes I^{\otimes N}} (|0\rangle + |1\rangle)/\sqrt{2} \otimes |0^{\otimes N}\rangle. \tag{22}$$

Step 3: Apply $C - Amp(\theta)$:

$$|\varphi\rangle \xrightarrow{C-Amp(\theta)} \\ (|0\rangle \otimes |0^{\otimes N}\rangle + |1\rangle \otimes Amp(\theta)|0^{\otimes N}\rangle)/\sqrt{2}. \tag{23}$$

Step 4: Add $C - Amp^\dagger(\theta')$:

$$|\varphi\rangle \xrightarrow{C-Amp^\dagger(\theta')} \\ (|0\rangle \otimes Amp^\dagger(\theta')|0^{\otimes N}\rangle + |1\rangle \otimes Amp(\theta)|0^{\otimes N}\rangle)/\sqrt{2}. \tag{24}$$

Step 5: Employ the Hadamard gate again:

$$|\varphi\rangle \xrightarrow{H \otimes I^{\otimes N}} [|0\rangle \otimes (Amp^\dagger(\theta') + Amp(\theta))|0^{\otimes N}\rangle \\ + |1\rangle \otimes (Amp^\dagger(\theta') - Amp(\theta))|0^{\otimes N}\rangle]/2. \tag{25}$$

In the end, according to Eq. (6), the quantum overlap is

$$\mathrm{Re}(\langle 0^{\otimes N}|Amp(\theta)Amp^\dagger(\theta')|0^{\otimes N}\rangle) = 2P(|0\rangle)_{j,k} - 1, \tag{26}$$

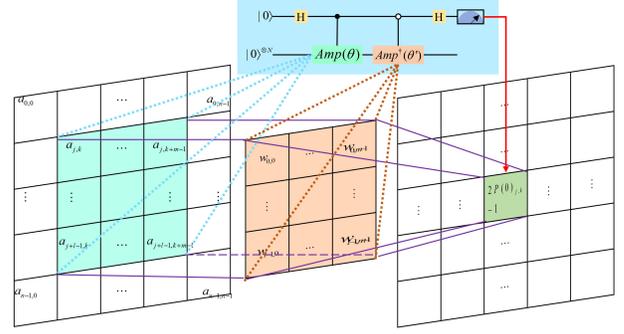

Fig. 3: A quantum adjoint convolution operation based on Hadamard test.

where $\langle 0^{\otimes N}|Amp(\theta) = \langle \Phi_{j,k}|$ and $Amp^\dagger(\theta')|0^{\otimes N}\rangle = |\Psi\rangle$.

Nevertheless, QACO in Fig. 3 accomplishes the Frobenius inner product computation only once, which hardly satisfies the growing demand. Though this ansatz can theoretically be executed repeatedly $l \times m$ times, it fails to embody the convenience of parallel processing. The ideal state is to construct a QACL based on QACO ansatz to solve all Frobenius inner product values. The structural similarity between QACO ansatz and QPE foreshadows the feasibility of this idea, implying that muti-phase estimation can extend this ansatz into a QACL. Subsequently, an elaboration is provided on how this can be achieved.

*2) Ansatz Structure of Quantum Adjoint Convolutional Layer:* Firstly, according to Eq. (25), Step 5 is reformulated as

$$|\varphi\rangle = [|0\rangle \otimes (|\chi\rangle + |\omega\rangle) + |1\rangle(|\chi\rangle - |\omega\rangle)]/2, \tag{27}$$

where $|\chi\rangle = |Amp^\dagger(\theta')0^{\otimes N}\rangle$. $|\omega\rangle = |Amp(\theta)0^{\otimes N}\rangle$. $P(|0\rangle)_{j,k} = [1 + \mathrm{Re}(\langle \chi|\omega\rangle)]/2$ is obtained from Eq. (26). Then $P(|1\rangle)_{j,k} = [1 - \mathrm{Re}(\langle \chi|\omega\rangle)]/2$. Let $\sin\theta = \sqrt{P(|0\rangle)_{j,k}}$ and $\cos\theta = \sqrt{P(|1\rangle)_{j,k}}$, then Eq. (27) is equivalent to

$$|\varphi\rangle = \sin\theta|0\rangle|\mu\rangle + \cos\theta|1\rangle|v\rangle, \tag{28}$$

where $|\mu\rangle = |\chi\rangle + |\omega\rangle$, $|v\rangle = |\chi\rangle - |\omega\rangle$, $\theta \in [0, \pi/2]$, $\cos 2\theta = -\mathrm{Re}(\langle \chi|\omega\rangle)$. The trigonometric functions in Eq. (28) also have the following connections to Euler formula, namely $\sin\theta = (e^{i\theta} - e^{-i\theta})/(2i)$ and $\cos\theta = (e^{i\theta} + e^{-i\theta})/2$. Thus Eq. (28) is rewritten as

$$|\varphi\rangle = -i(e^{i\theta}|\Phi^+\rangle - e^{-i\theta}|\Phi^-\rangle)/\sqrt{2}, \tag{29}$$

where $|\Phi^\pm\rangle = (|0\rangle|\mu\rangle \pm i|1\rangle|v\rangle)/\sqrt{2}$. At this point, a

$$U = (I^{\otimes(N+1)} - 2|\varphi\rangle\langle\varphi|)(Z \otimes I^{\otimes N}) \tag{30}$$

can be constructed to evaluate the eigenvalue $\pm\theta_\varphi$ in

$$\begin{aligned} U|\Phi^\pm\rangle &= (I^{\otimes(N+1)} - 2|\varphi\rangle\langle\varphi|)|\Phi^\mp\rangle \\ &= \left[-2\frac{-i}{\sqrt{2}}(e^{i\theta}|\Phi^+\rangle - e^{-i\theta}|\Phi^-\rangle) \right. \\ &\quad \left. \times \frac{i}{\sqrt{2}}(e^{-i\theta}|\Phi^+\rangle - e^{i\theta}|\Phi^-\rangle) + I^{\otimes(N+1)}\right]|\Phi^\mp\rangle \\ &= e^{\pm 2i\theta}|\Phi^\pm\rangle, \end{aligned} \tag{31}$$

where $Z$ denotes the Pauli $Z$ operator. Since $|\varphi\rangle$ is viewed as evolving from an unitary operator $U_\varphi$, then Eq. (30) is re-expressed as

$$U = U_\varphi \mathcal{G} U_\varphi^\dagger (Z \otimes I^{\otimes N}) \qquad (32)$$

in Fig. 4, where

$$\mathcal{G} = I^{\otimes(N+1)} - 2|0^{\otimes(N+1)}\rangle\langle 0^{\otimes(N+1)}| \qquad (33)$$

is shown in Fig. 5.

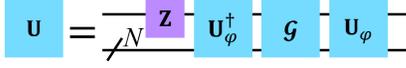

Fig. 4: Quantum circuits of $U$.

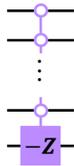

Fig. 5: Quantum circuits of $\mathcal{G}$.

Next, multi-QPE is applied. Combining Eq. (29) and Eq. (31), the approximate solution of QPE can be obtained as

$$|\tilde{\varphi}\rangle = -i(e^{i\theta}|y\rangle|\Phi^+\rangle - e^{-i\theta}|2^s - y\rangle|\Phi^-\rangle)/\sqrt{2}, \qquad (34)$$

where $y \in [0, 2^{s-1}]$, $y\pi/2^{s-1} \approx 2\theta$. Eventually, **QACL ansatz**

$$[QFT^{\dagger \otimes(k-1)s} \otimes I^{\otimes(N+1)}] \prod_i C - U_i \times \\ [H^{\otimes(k-1)s} \otimes I^{\otimes(N+1)}]|0^{\otimes((k-1)s+N+1)}\rangle \qquad (35)$$

with multi-QPE in Fig. 6 is devised by us to calculate all the eigenvalues $\{|\tilde{\varphi}\rangle_i\}_{i=1}^k$, where $C - U_i$ denotes a controlled $U$ in Eq. (32). $\{|\tilde{\varphi}\rangle_i\}_{i=1}^k$ are used to represent all the Frobenius inner product results, where $k$ is related to strides. The individual QPE solution processes in Eq. (35) are explained by Eq. (27) to Eq. (34). Thus the multi-QPE process is a parallel execution of the single QPE solving process and is not described here.

In summary, in this subsection, the ansatz structure of QACO as well as QACL is designed. The link between QACO and QACL is established by QPE. In essence, QACL exhibits interpretability and data characterization efficiency, while concurrently enabling parallel computation of all Frobenius inner products owing to its ansatz structure.

## IV. EXPERIMENTS

In this section, the novelty and effectiveness of QACO and QACL are manifested in the following three experiments under noiseless environment on PennyLane and TensorFlow platforms:

- The validity and correctness are verified by a numerical experiment that counts 10, 100, 1000 and 10000 measurements of QACO.
- For the same image, the differences between QACL and the other convolution operations are compared when the convolution kernel parameters are fixed.
- The performance of QACL is tested on the MNIST and Fashion MNIST datasets when the parameters in the kernel are trainable.

Simultaneously, it is imperative to acknowledge that QACL cannot be deployed on free publicly available quantum computers with limited hardware resources due to its deeper circuit depth and more qubits. For example, the current freely accessible quantum computer from IBM offers only five to seven qubits. Therefore, numerical and quantum simulations are mainly adopted here. Moreover, computational acceleration is achieved by quantum device "lightning.qubit" in PennyLane and two 2080 Ti GPUs.

### A. Verification of Quantum Adjoint Convolutional Operation

Here, a numerical experiment is presented herein to substantiate the efficacy of QACO, elucidating differentials between actual and theoretical outcomes. First, assuming a 3-qubit system with two given quantum states $\langle\Phi_{j,k}| = [1/6, 1/6, -5/6, 3/6]$ and $|\Psi\rangle = [5/6, 3/6, 1/6, 1/6]^T$, the precise solution, as derived from Eq. (9), is $1/6$. Subsequently, these quantum states are encoded into two controlled ansatzes $(C - Amp(\theta)$ and $C - Amp^\dagger(\theta'))$ in Fig. 7. The outcomes of performing 10, 100, 1000, and 10000 measurements of $|0\rangle$ are detailed in Tab. I. The tabulated data illustrates that with an increasing number of measurements, the statistical

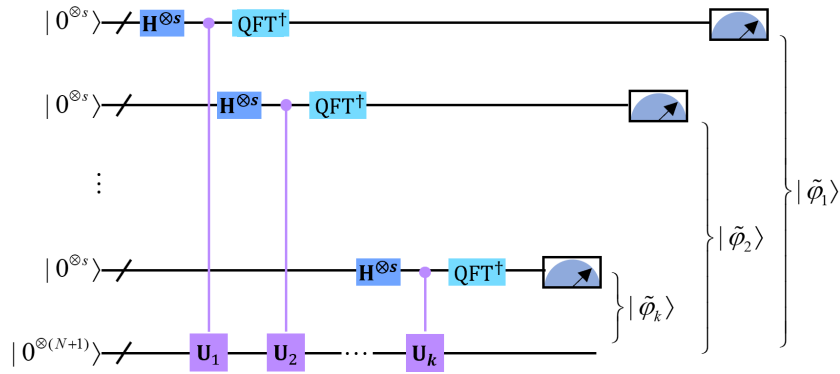

Fig. 6: Quantum adjoint convolutional layer based on quantum phase estimation.

results converge towards the theoretically expected value of $1/6$. The observed fluctuations in error could be attributed to (1) truncation errors arising during the classical-to-ansatz data mapping and (2) an insufficient number of statistical

Tab. I: Comparison between Eq. (9) and actual values

| number of measurements | measurement probability for $|0\rangle$ | measurement error |
|---|---|---|
| 10 | 0.6 | 16.67% |
| 100 | 0.58 | 4.17% |
| 1000 | 0.573 | 14.16% |
| 10000 | 0.582 | 1.63% |

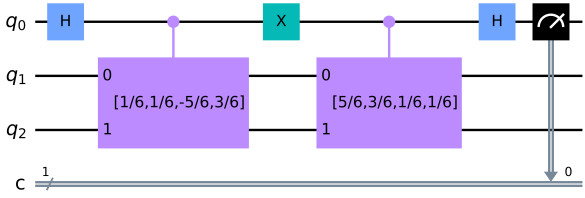

Fig. 7: 3-qubit system.

measurements leading to unstable outcomes.

### B. Quantum Adjoint Convolutional Layer with Fixed Parameters

When the parameters of kernel $C - Amp^{\dagger}(\theta')$ in QACL are held constant, a comparative analysis is conducted among QACL, adjoint method, swap test, and classical convolution operation. Here, five typical kernels are selected as shown in Tab. II, including edge detection, Gaussian blur, etc. In addition, as the original image, a three-channel $50 \times 50$ image from the Internet, would consume a large number of qubits, a numerical simulation of Eq. (35) is performed by PennyLane when padding is 0 and stride is set to 1 and 2. The corresponding outcomes are depicted in Fig. 8. Commencing from the second column, each column in Fig. 8 denotes a distinct predetermined kernel, while each horizontal axis corresponds to a convolution operation. The visualizations in Fig. 8 unveil that the quantum schemes, namely adjoint method, swap test, and QACL, elicits disparate insights for an identical original image, when all convolution kernel parameters remain uniform. In contrast, QACL and swap test provide a richer understanding, while the adjoint method is relatively monotonous. Although these various insights are not entirely consistent with the classical convolution operation, it does not mean that QACL is less effective than classical. Keep in mind that QACL is not a strict imitator of CCLs. It provides a new perspective, which causes its results not to be reasonable to interpret from a classical machine learning perspective. Therefore, this divergence is also seen as an interesting innovation. Albeit, direct comparison between classical convolutional operations and quantum convolution is hard, the source of this difference can be found between three comparable quantum methods. The convolution results of the three quantum methods are relatively close to each other, which is still rooted in the structural differences in the ansatz that lead to different measured quantum overlaps.

Tab. II: Convolution operation

| name | matrix representation |
|---|---|
| edge detection | $\begin{bmatrix} -1 & -1 & -1 \\ -1 & 8 & -1 \\ -1 & -1 & -1 \end{bmatrix}$ |
| Gaussian blur | $\frac{1}{16}\begin{bmatrix} 1 & 2 & 1 \\ 2 & 4 & 2 \\ 1 & 2 & 1 \end{bmatrix}$ |
| sharpen | $\begin{bmatrix} 0 & -1 & 0 \\ -1 & 5 & -1 \\ 0 & -1 & 0 \end{bmatrix}$ |
| emboss | $\begin{bmatrix} -2 & -1 & 0 \\ -1 & 1 & 1 \\ 0 & 1 & 2 \end{bmatrix}$ |
| box blur | $\frac{1}{9}\begin{bmatrix} 1 & 1 & 1 \\ 1 & 1 & 1 \\ 1 & 1 & 1 \end{bmatrix}$ |

### C. Quantum Adjoint Convolutional Layer with Trainable Parameters

Here, the parameters in the convolutional kernel in QACL that can be iteratively updated are considered. Subsequently, QACL is evaluated for performance on both MNIST and Fashion MNIST datasets.

*1) Dataset:* MNIST [32] and Fashion MNIST [33] stand as entrenched and widely embraced benchmark datasets in the domain of machine learning, constituting foundational elements for myriad studies in the field. Each dataset comprises 10,000 test images and 60,000 training images, with dimensions of $28 \times 28$ pixel points. In our experimental setup, 550 images labeled 0 and an additional 550 with label 1 are selected from MNIST (or Fashion MNIST). Within each label subset, 500 images are randomly allocated for the training set, while the remainder serve as the test set. Here, in order to ensure the smooth implementation of QACL, the original image size is reduced to $14 \times 14$ by the principal component analysis algorithm.

*2) Comparison of models:* For a CCNN, its network structure is indicated as

$$\text{Cov}_1 \circ \text{Pool}_1 \circ \text{Cov}_2 \circ \text{Pool}_2 \circ \text{FC}_1 \circ \text{FC}_2, \quad (36)$$

where $\text{Cov}_i$ is CCL. $\text{Pool}_i$ is a pooling layer. $\text{FC}_i$ is a fully connected layer. $\circ$ denotes a layer-to-layer connection. In this section the first convolutional layer $Cov_1$ is replaced by QACL, i.e.

Tab. III: Average precision of the last 10 epochs

| models | average precision |
|---|---|
| model (36) for MNIST | 0.881 |
| model (36) for Fashion MNIST | 0.784 |
| model (37) for MNIST | 0.964 |
| model (37) for Fashion MNIST | 0.846 |

$$QACL \circ Pool_1 \circ Cov_2 \circ Pool_2 \circ FC_1 \circ FC_2. \qquad (37)$$

Besides, some details are highlighted here. Here, the activation function and the size of the kernel are uniformly set to ReLU and $4 \times 4$, respectively. $Cov_1$ and $Cov_2$ have 50 and 64 kernels respectively. $FC_1$ has 1024 neurons and the parameter of the dropout layer is 0.5. $FC_2$ contains 10 neurons. QACL is mathematically modeled according to Eq. (35).

*3) Experimental results:* The comparison outcomes of model (36) and (37) is shown in Fig. 9. In Fig. 9, the horizontal coordinate denotes epoch. The vertical coordinates refer to test accuracy and loss, respectively. The blue and orange (or green and red) lines represent model (36) and model (37) for MNIST (or Fashion MNIST). The comparison leads to the following conclusions. (1) By counting the average accuracies of the last 10 epochs, as shown in Tab. III, it is found that the QACL improves the accuracy by at least 0.6, which is a significant feature, no matter which dataset is used. (2) The test accuracy of QACL on MNIST is about 0.12 higher than that on Fashion MNIST, which demonstrates that it is more effective on the MNIST dataset. (3) Analysis of the loss shows that the loss value without using QACL is lower, indicating that model (36) has stronger learning performance. (4) Taken together, QACL significantly improves test accuracy to some extent at the cost of learning performance.

## V. Conclusion

To solve the problems of incomprehensible black-box design and inefficient data encoding in QCLs, this paper theoretically shows that the principle of QACO is equivalent to the quantum normalization result of Frobenius inner product. Specifically, the Frobenius inner product can be realized by Hadamard test and two controlled quantum amplitude encodings. When QACO is combined with the QPE method, it can be extended to QACL, which facilitates solving all Frobenius inner products at once. To estimate the performance of QACL, a categorical discussion focusing on fixed and unfixed parameters of the kernel in QACL is presented. When the parameters of the kernel are fixed, QACL can provide insights into convolutional operations with a quantum perspective that is not necessarily identical to classical convolutional operations. When the parameters in the kernel are trainable, classification experiments with MNIST and Fashion MNIST on the PennyLane and TensorFlow platforms show that QACL provides higher accuracy in classification tests, but at the expense of learning performance. In short, our research designs interpretable QCLs while taking into account the efficiency of their data representation, which can lay the foundation for various aspects of quantum-enhanced machine learning as well as quantum machine vision.

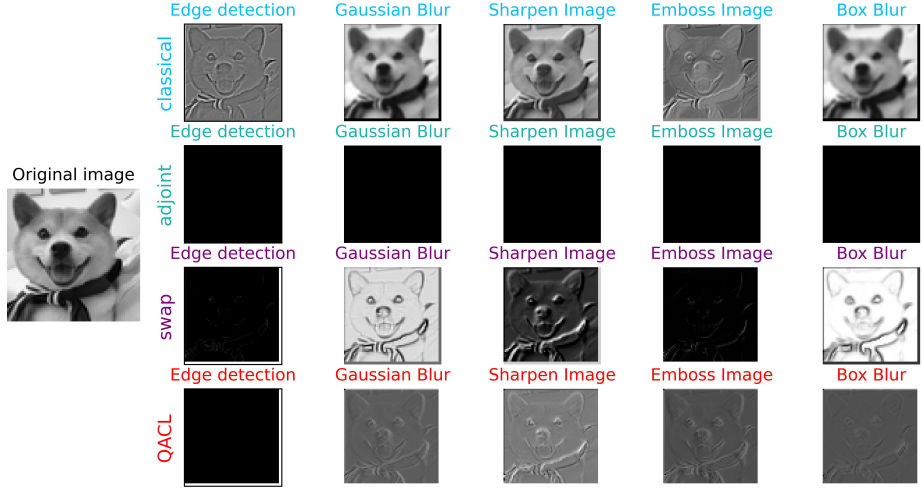

(a) Stride = 1.

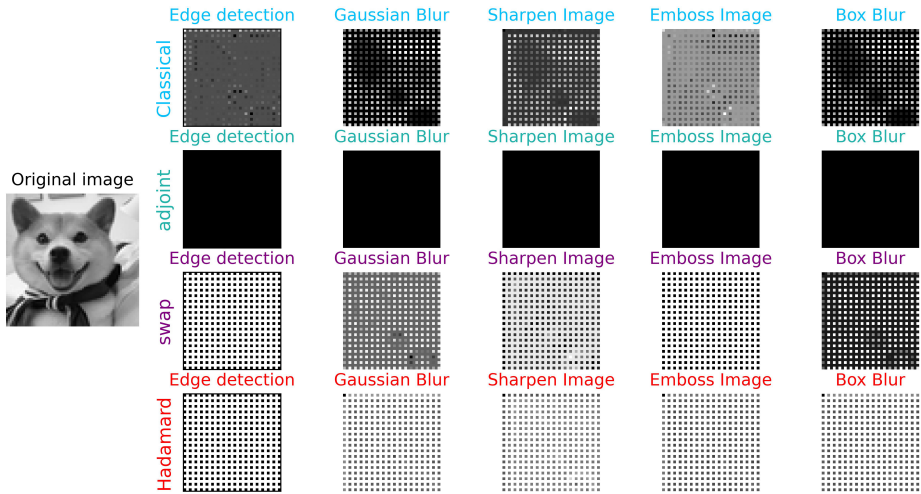

(b) Stride = 2.

Fig. 8: Comparison results.

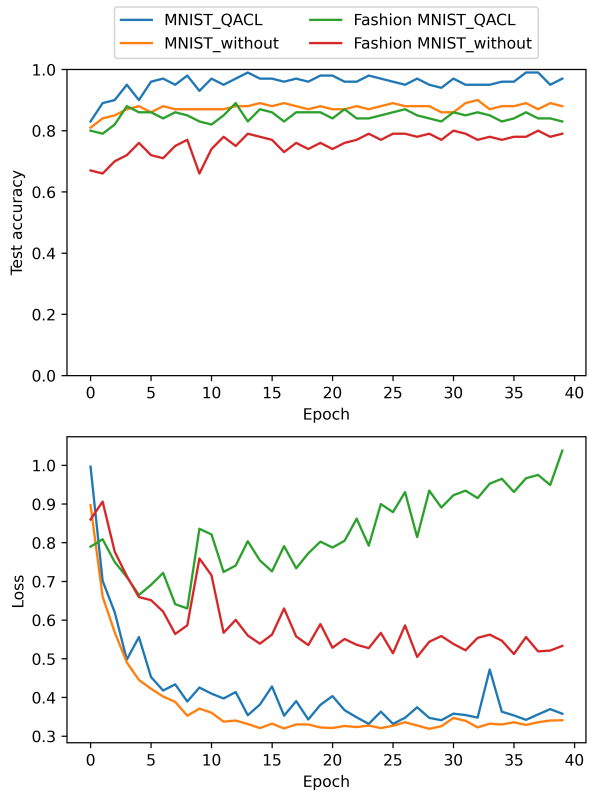

Fig. 9: Comparison results between model (36) and (37).